\documentclass[prl,twocolumn,superscriptaddress,showpacs]{revtex4}
\usepackage{graphicx}
\usepackage{amsmath}
\usepackage{amssymb}
\usepackage{dcolumn}
\usepackage{bm}
\usepackage{textcomp}
\usepackage{ulem}
\usepackage{colordvi}

\begin{document}

\title{Nuclear Charge Radii of $^{9,11}$Li: the Influence of Halo Neutrons}

\author{R.~S\'anchez}
\affiliation {Gesellschaft f\"ur Schwerionenforschung, 64291 Darmstadt, Germany}

\author{W.~N\"ortersh\"auser}
\affiliation {Gesellschaft f\"ur Schwerionenforschung, 64291 Darmstadt, Germany} %
\affiliation {Departement of Physics, University T\"ubingen, D-72076 T\"ubingen, Germany}

\author{G.~Ewald}
\affiliation {Gesellschaft f\"ur Schwerionenforschung, 64291 Darmstadt, Germany}

\author{D.~Albers}
\affiliation {TRIUMF, Vancouver, BC, Canada V6T 2A3}

\author{J.~Behr}
\affiliation {TRIUMF, Vancouver, BC, Canada V6T 2A3}

\author{P.~Bricault}
\affiliation {TRIUMF, Vancouver, BC, Canada V6T 2A3}

\author{B.A.~Bushaw}
\affiliation {Pacific Northwest National Laboratory, P.O. Box 999, Richland, WA 99352, USA}

\author{A.~Dax}
\altaffiliation[Current address ]{CERN, CH-1211 Geneva 23, Switzerland.} %
\affiliation {Gesellschaft f\"ur Schwerionenforschung, 64291 Darmstadt, Germany}

\author{J.~Dilling}
\affiliation {TRIUMF, Vancouver, BC, Canada V6T 2A3}

\author{M.~Dombsky}
\affiliation {TRIUMF, Vancouver, BC, Canada V6T 2A3}

\author{G.W.F.~Drake}
\affiliation {Department of Physics, University of Windsor, Windsor, Ontario, Canada, N9B 3P4}

\author{S.~G\"otte}
\affiliation {Gesellschaft f\"ur Schwerionenforschung, 64291 Darmstadt, Germany}

\author{R.~Kirchner}
\affiliation {Gesellschaft f\"ur Schwerionenforschung, 64291 Darmstadt, Germany}

\author{H.-J.~Kluge}
\affiliation {Gesellschaft f\"ur Schwerionenforschung, 64291 Darmstadt, Germany}

\author{Th.~K\"uhl}
\affiliation {Gesellschaft f\"ur Schwerionenforschung, 64291 Darmstadt, Germany}

\author{J.~Lassen}
\affiliation {TRIUMF, Vancouver, BC, Canada V6T 2A3}

\author{C.D.P.~Levy}
\affiliation {TRIUMF, Vancouver, BC, Canada V6T 2A3}

\author{M.R.~Pearson}
\affiliation {TRIUMF, Vancouver, BC, Canada V6T 2A3}

\author{E.J.~Prime}
\affiliation {TRIUMF, Vancouver, BC, Canada V6T 2A3}

\author{V.~Ryjkov}
\affiliation {TRIUMF, Vancouver, BC, Canada V6T 2A3}

\author{A.~Wojtaszek}
\altaffiliation[Current address ]{Institute of
Physics, Swietokrzyska Academy, PL-25-406, Kielce, Poland.}
\affiliation {Gesellschaft f\"ur
Schwerionenforschung, 64291 Darmstadt, Germany}

\author{Z.-C.~Yan}
\affiliation {Department of Physics, University of New Brunswick, Fredericton, New Brunswick,
Canada E3B 5A3}

\author{C. Zimmermann} \affiliation {Departement of Physics, University T\"ubingen,
D-72076 T\"ubingen, Germany}

\date{\today}

\begin{abstract}
The nuclear charge radius of $^{11}$Li has been determined for the
first time by high precision laser spectroscopy. On-line
measurements at TRIUMF-ISAC yielded a $\rm ^7Li - {^{11}Li}$ isotope
shift (IS) of 25\,101.23(13) MHz for the Doppler-free
$2s\;{^2S_{1/2}} \to 3s\;{^2S_{1/2}}$ transition. IS accuracy for
all other bound Li isotopes was also improved. Differences from calculated mass-based IS yield values for change
in charge radius along the isotope chain. The charge radius decreases monotonically from $^6$Li to $^9$Li, and
then increases from 2.217(35) fm to 2.467(37) fm for $^{11}$Li. This is compared to various models, and it is
found that a combination of halo neutron correlation and intrinsic core excitation best reproduces the
experimental results.
\end{abstract}
\maketitle

For twenty years, halo nuclei with diffuse outer neutron distributions have been known to exist at the limits of
stability for many of the lighter elements \cite{Jen04}. The first discovered \cite{Tan85} and most renowned of
these is $^{11}$Li with two halo neutrons; however, details of the nuclear structure and halo -- core
interactions are still not well understood. Nuclear forces are not strong enough to bind a neutron to $^9$Li,
nor can they bind two neutrons into a dineutron. Yet adding two neutrons to $^9$Li leads to a bound nucleus $-$
$^{11}$Li ($T_{1/2}$ = 8.4 ms), illustrating the importance of understanding the interaction that allow formation
of the halo structure. Recent measurement \cite{Wan04} of the {\it rms} nuclear charge radius ($r_{\rm c}$) for
the two-neutron halo $^6$He indicates that its halo is a dineutron "orbiting" the $^4$He core. The core is a
strongly-bound $\alpha$-particle and model calculations \cite{Pie01} estimate only a 4\% increase in $r_{\rm
c}(\alpha)$. In contrast, the $^9$Li-like core of $^{11}$Li is 'softer' and interaction between halo neutrons
and core nucleons may significantly polarize the core.

An indicator for an altered $^9$Li core would be a change in proton distribution between $^9$Li and $^{11}$Li.
This was investigated in collisions which removed a proton from a $^{11}$Li projectile \cite{Bla92}, but within
the rather large uncertainty, there was no clear evidence for a change in the deduced charge radius. Also,
analysis and interpretation was not straightforward because of the dependence on an assumed nuclear model. A
more sensitive approach to determine the change in $r_{\rm c}$ is a measurement of the isotope shift in an
atomic transition \cite{Yan00}. A finite nuclear charge distribution reduces electron binding energies,
particularly for $s$-electrons that have probability of being inside the nucleus, and a change in the
distribution between isotopes can be observed as shifts in electronic transition energies. In light elements,
the mass-based isotope shift is much larger than the nuclear volume shift; for lithium, about 10,000 times
larger. The dominant portion of the mass shift is change in reduced mass (normal mass shift), but electron
correlations (specific mass shift) are also important. Recent high-precision calculations account for these
correlations, as well as relativistic and QED corrections \cite{Yan03}. In this work we present the first
measurement of the $^{11}$Li isotope shift in the $2s \to 3s$ Doppler-free two-photon transition, as well as
refined values for all other isotopes. These are compared with calculated mass shifts, yielding nuclear charge
radii that are compared with various theoretical models and interpreted in terms of halo correlation and core
polarization.

The experiment must fulfil two conditions: measure the isotope
shift to an uncertainty of one part in $10^5$ ($10^{-10}$ of total
transition frequency), and provide an overall efficiency
sufficient to observe the resonances with production yields of
$\sim 10^4$ $^{11}$Li atoms/s. Moreover, the short half-life
requires on-line study at the production facility. We previously
reported a technique \cite{Ewa04} to perform such measurements and
used it to determine the charge radii of $^{8,9}$Li produced at
the GSI-UNILAC. For the experiments reported here, the apparatus
was moved to the TRIUMF-ISAC facility in Vancouver, Canada where
$^{11}$Li is produced by a 40 \textmu\@A, 500 MeV proton beam
impinging on a tantalum target. $^{11}$Li$^+$ ions extracted from
the target (typ. 30,000/s) are implanted in a hot carbon foil
where they are neutralized and released as atoms into the
low-field source region of a quadrupole mass spectrometer (QMS).
The neutral atoms are re-ionized via doubly-resonant four-photon
ionization:

\begin{equation}
2s\xrightarrow{\rm 2\times 735nm}3s\xrightarrow[{\rm 30 ns}]{\rm decay}2p\xrightarrow{\rm
610nm}3d\xrightarrow{\rm 610,735nm}{\rm Li^+}~,
\end{equation}
with excitation taking place at the center of a doubly-resonant
optical enhancement cavity ($\sim$$100\times$) built around the
QMS source region. The titanium-sapphire laser that excites the
$2s \to 3s$ two-photon transition is beat-frequency servo-locked
to an I$_2$ hyperfine line stabilized diode laser. As previously described \cite{Ewa04}, measurements on
$^{11}$Li (and the other isotopes) were interspersed with measurements on $^6$Li which served as the experimental
reference, and measured optical powers were used to correct for calibrated AC-Stark shifts.

\begin{figure}[tbp]
\includegraphics [scale=0.7]{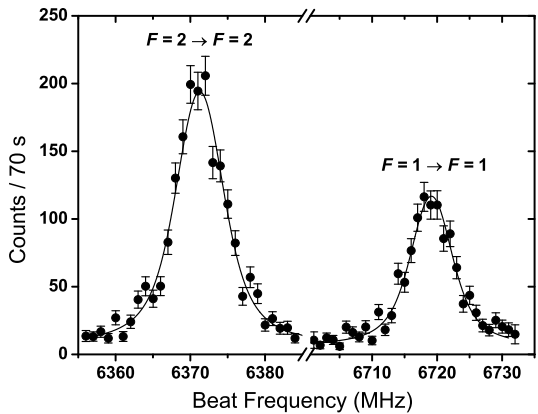}
\caption{Resonances in the $2s \to 3s$ transition of $^{11}$Li as
a function of the beat frequency between the titanium-sapphire
laser and the reference diode laser. Error bars are simple
counting statistics on the number of observed ion counts. }
\label{fig:fig1}
\end{figure}
Figure\,\ref{fig:fig1} shows a typical $^{11}$Li spectrum.
Twenty-four such spectra were obtained over six days of beam time.
With nuclear spin $I = 3/2$, the $^2S_{1/2}$ states have $F =
1,~2$ hyperfine components, which obey the two-photon selection
rule  $\Delta F = 0$ for an $s \to s$ transition. All Li isotopes
have nuclear spin and exhibit similar doublets: Isotope shifts are
taken with respect to center-of-gravities of the two hyperfine
lines for each isotope. Results for all isotopes, relative to
$^7$Li, are given in Table\,\ref{tab:table1}. Values for
$^{6,8,9}$Li are in good agreement with our previous measurements
\cite{Ewa04}, but with improved precision. The $^6$Li isotope shift was also determined earlier with a different
technique as -11\,453\,734(30)~kHz \cite{Bus03}; this is significantly different from our current measurements
($\sim 5$ times the combined uncertainties), and is attributed to unaccounted systematic errors in the prior
interferometric measurements \cite{Bus03}, as compared to the current frequency-based measurements. The IS for
the halo nucleus $^{11}$Li is a first-time measurement.

Successful determination of changes in $r_{\rm c}$ from the isotope shift
measurements depends critically on the combined accuracy of theory and experiment.
\begin{table}
\caption{\label{tab:table1} Isotope shifts  measured at TRIUMF (this
work) and GSI \cite{Ewa04} [avg = weighted mean] compared with theoretical mass shifts  for $\rm ^7Li - {^ALi}$
in the
$2s\;{^2S_{1/2}} \to 3s\;{^2S_{1/2}}$ transition. Uncertainties for
$r_{\rm c}$ are dominated by uncertainty in the reference radius
$r_{\rm c}(^7{\rm Li}) = 2.39(3)~{\rm fm}$ \cite{Jag74}.}
\begin{ruledtabular}
\begin{tabular}{r D{,}{\,}{12} D{,}{\,}{12} D{.}{.}{6}}
Isotope & \multicolumn{1}{c}{Isotope Shift, kHz} & \multicolumn{1}{c}{Mass Shift, kHz} & \multicolumn{1}{c}{$r_{\rm c}$\,, fm}\\
\hline
$^6$Li~~~this & -11,453\,984(20) \\
         GSI  & -11,453\,950(130) \\
         avg  & -11,453\,983(20) & -11,453\,010(56) & 2.517(30)\\
$^8$Li~~~this &   8,635\,781(46) \\
         GSI  &   8,635\,790(150)\\
         avg  &   8,635\,782(44) &   8,635\,113(42) & 2.299(32)\\
$^9$Li~~~this &  15,333\,279(40) \\
         GSI  &  15,333\,140(180)\\
         avg  &  15,333\,272(39) &  15,332\,025(75) & 2.217(35)\\
$^{11}$Li~~~this & 25,101\,226(125)^{\rm a} & 25,101\,812(123) & 2.467(37)\\
\end{tabular}
\end{ruledtabular}
$^{\rm a}$ 68 kHz statistical + 57 kHz systematic from AC-Stark
shift
\end{table}
On the theoretical side, the quantum mechanical many-body problem
must be solved to high accuracy in the nonrelativistic limit, and
then the effects of relativity and quantum electrodynamics are
included with perturbation theory.  In the past, theoretical
results with laser-spectroscopic accuracy were not available for
atoms more complicated than helium, even in the nonrelativistic
limit. This problem is now solved by variational methods involving
correlated basis sets with multiple distance scales \cite{Yan00}.
The resulting electron wave functions are used to calculate the
various contributions to the mass shift, listed for $^{7,11}$Li in
Table\,\ref{tab:table2}. A recent first calculation \cite{Yan03}
of the mass polarization correction to the Bethe logarithm part of
the electron self-energy has significantly reduced uncertainty in
the QED contribution; overall calculation uncertainty is now
limited by the  relativistic recoil term of order
$\alpha^2(\mu/M)$.
\begin{table}[b]
\caption{\label{tab:table2} Contributions to the $\rm ^7Li -
{^{11}Li}$ mass shift in the $2s\;{^2S_{1/2}} \to 3s\;{^2S_{1/2}}$
transition, excluding nuclear size effects. $\mu/M$ is the ratio
of the reduced mass and the atomic mass. Uncertainty in the
nonrelativistic $(\mu/M)$ term is from uncertainty in the
$^{11}$Li mass \cite{Bac05}, while limiting uncertainty in the
relativistic and QED terms is computational.}
\begin{ruledtabular}
\begin{tabular}{l D{,}{\,}{5}}
Contribution (order) & \multicolumn{1}{c}{kHz} \\
\hline
Nonrelativistic $(\mu/M)$     & 25\,104\,483,(20)\\
Nonrelativistic $(\mu/M)^2$   &      -2\,968,(0)\\
Relativistic $\alpha^2(\mu/M)$&          417,(121)\\
QED $\alpha^3(\mu/M)$         &         -120,(6)\\
Total                         & 25\,101\,812,(123)\\
\end{tabular}
\end{ruledtabular}
\end{table}

The total in Table\,\ref{tab:table2} is the calculated mass-based
component of the isotope shift; corresponding shifts for all
isotopes are obtained directly from coefficients given in Table~III
of Ref. \cite{Yan03} and are listed in Table\,\ref{tab:table1}.
Differences from measured isotope shifts are then attributed
to the nuclear volume
effect and are related to $r_{\rm c}$ of the two isotopes by
\begin{multline}
 \delta\nu_{\rm IS,exp}^{A,7}-\delta\nu_{\rm IS,MS}^{A,7}\\
=\frac{Ze^2}{3\hbar}\left[r_{\rm c}^2({^A{\rm Li}})-r_{\rm
c}^2({^7{\rm
Li}})\right]\left(\left\langle\delta(r_i)\right\rangle_{3s}-
\left\langle\delta(r_i)\right\rangle_{2s}\right)\\
= -1.5661\frac{\rm MHz}{\rm fm^2}\left[r_{\rm c}^2({^A{\rm
Li}})-r_{\rm c}^2({^7{\rm Li}})\right]~,
\end{multline}
where $Ze$ is the nuclear charge and
$\left\langle\delta(r_i)\right\rangle$ are expectation values for
electron density at the nucleus  in the respective states
\cite{Yan00}.

Optical isotope shift measurements provide only the change in the
{\it rms} nuclear charge radius between two isotopes. Absolute
charge radii $r_{\rm c}$ must be referenced to at least one
isotope that is determined with a different technique. For the
stable $^{6,7}$Li isotopes, $r_{\rm c}$ have been determined by
elastic electron scattering \cite{Jag74}, from which we use
$r_{\rm c}(^7{\rm Li})=2.39(3)$ fm as a reference radius. This and
the measured $^{6,7}$Li isotope shift yields $r_{\rm c}(^6{\rm
Li})=2.52(3)$ fm, in good agreement with the electron scattering
result of 2.55(4) fm \cite{Jag74}. Combining measured isotope
shifts, calculated mass shifts, and the $^7$Li reference radius
yields $r_{\rm c}$ for the other isotopes, as given in the last
column of Table\,\ref{tab:table1}.

\begin{figure}[tbp]
\includegraphics [scale=0.7]{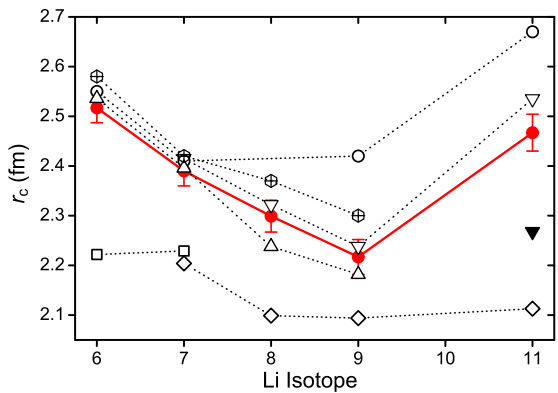}
\caption{(Color online) Experimental charge radii of lithium
isotopes (\Red{$\bullet$}) compared with theoretical predictions: {\footnotesize $\triangle$}: Greens-Function
Monte Carlo Calculations \cite{Pie02, Pie01}, $\triangledown$: Stochastic Variational Multi-Cluster Model
\cite{Var95, Var02} ($\blacktriangledown$: assuming a frozen $^9$Li core), $\oplus$: Fermionic Molecular
Dynamics \cite{Nef05},  $\circ$: Dynamic Correlation Model \cite{Tom01}, $\Box$ and $\diamond$: ab-initio
No-Core Shell Model \cite{Nav03, Nav98}.} \label{fig:fig2}
\end{figure}
The derived nuclear charge radii are shown as filled circles in
Fig.\,\ref{fig:fig2}: while $r_{\rm c}$ decreases continuously
from $^6$Li to $^9$Li, there is a large increase from $^9$Li to
$^{11}$Li. The significance of these results becomes evident when
compared with predictions from different nuclear models, also
shown in Fig.\,\ref{fig:fig2}. Models using point-proton radii
$r_{pp}$ are converted to nuclear charge radii $r_{\rm c}$ by
folding in proton \cite{Sic03} and neutron \cite{Kop97}
mean-square charge radii:
\begin{equation}
\left\langle r_{\rm c}^2\right\rangle = \left\langle r_{\rm
pp}^2\right\rangle + \left\langle R_{\rm p}^2\right\rangle +
\frac{N}{Z} \left\langle R_{\rm n}^2\right\rangle + \frac{3\hbar^2}{4m_{\rm
p}^2c^2}~,
\end{equation}
where the last term is the Darwin-Foldy correction for
``Zitterbewegung" of the proton \cite{Fri97}.

Neither conventional shell model nor self-consistent Hartree-Fock
calculations have correctly reproduced halo-specific anomalous
properties of light nuclei close to the neutron drip-line. Early
models for $^{11}$Li only treated its three-body character,
without considering possible polarization of the $^9$Li core
\cite{Zhu93}; thus, change in nuclear charge radius could only be
caused by correlation of the two halo neutrons. If they spend most
of their time on the same side of the core, the center-of-mass
(CM) is clearly different from the core center, the $^9$Li core
orbits the CM, and the averaged charge distribution is diffused.
Forss\'en {\it et al.} \cite{For02} constructed corresponding wave
functions for $^{11}$Li to obtain an analytical model for
electromagnetic dissociation of halo nuclei. The $^{11}$Li {\it
rms} matter radius of 3.55 fm was adjusted to be in good agreement
with experiment \cite{Ege02}; the predicted CM - core distance
$R_{\rm CM}$ ranged from 0.8 fm to 1.08 fm. The approximation
\cite{Zhu05} $r_{\rm c}(^{11}{\rm Li})=\left[R_{\rm CM}^2+r_{\rm
c}^2(^9{\rm Li})\right]^{1/2}= 2.40(6)$ fm is in reasonable
agreement with, but slightly lower than, our experimental result.
However, information available on the binary neutron-$^9$Li (core)
interaction is insufficient for these calculations to yield
structural details on $^{11}$Li, nor do they make predictions for
changes in $r_c$ between the non-halo nuclei.

The dynamic correlation model (DCM) is a more advanced scheme that starts from shell model states, and then
introduces neutron-core interaction with a two-body potential \cite{Tom01, Las62}. This leads to an admixture of
virtually excited single-particle states from the core. For $^{11}$Li, excited bound and continuum states of
$^9$Li up to 50 MeV were included in the analysis. Charge radii calculated ($\circ$ in
Fig.\,\ref{fig:fig2}) for $^{6,7}$Li agree well with our measurements, and while those for $^{9,11}$Li are
clearly overestimated, the increase from $^9$Li to $^{11}$Li is correctly reflected.

More sophisticated nuclear models treat interactions between individual nucleons using realistic nucleon-nucleon
({\it NN}) and three-nucleon ({\it NNN}) interactions. {\it NN} potentials are usually based on the multi-energy
partial-wave analysis of elastic {\it NN} scattering data produced by the Nijmegen group \cite{Sto93} in 1993,
while the {\it NNN} interaction parameters are adjusted to fit the binding energies of light nuclei.
Greens-Function Monte-Carlo (GFMC) calculations \cite{Pie02, Pie01}, the most fundamental description available
for light nuclei, have been completed for most nuclei with mass numbers $A \leq 12$. Results for the isotopes
$^{6,7,8,9}$Li are shown ($\triangle$) in Fig.\,\ref{fig:fig2} and are in good agreement with the experimental
results. The general trend is reproduced, but thus far the model has not been able to reproduce the $^{11}$Li
binding energy.


No-Core Shell Model (NCSM) calculations have been performed using
realistic {\it NN} potentials. Earlier calculations \cite{Nav98}
($\diamond$) for $^{7,8,9,11}$Li treated three-body interactions
as an effective phenomenological potential, while recent work
\cite{Nav03} ($\Box$) included microscopic three-body potentials
and was applied to $^{6,7}$Li. As seen in Fig.\,\ref{fig:fig2},
neither the absolute charge radii, nor the trend along the
isotopic chain are in agreement with our results.

The Fermionic Molecular Dynamics (FMD) model \cite{Nef05b} uses Gaussian wave packets for individual nucleons.
The {\it NN}-interaction is derived from the Argonne V18 interaction, treating short-range correlations
explicitly with a unitary operator. Predictions of the model \cite{Nef05} ($\oplus$) are in good agreement with
experiment for $^{6,7,8,9}$Li, but, like GFMC, the halo structure of $^{11}$Li has not yet been successfully
modelled.

Calculations that consider interactions between all individual
nucleons quickly become very complex and time consuming with
increasing nucleon number. Cluster models like the stochastic
variational multi-cluster (SVMC) calculations of Varga {\it et
al.} \cite{Var95, Var02} freeze some parts of the model space and
allow focus on those degrees of freedom thought to be most
relevant to the physical behaviour of a given nucleus. To a large
extent, this cluster structure can also be identified in FMD
calculations. The building blocks in the SVMC model are the
nucleons $p$ and $n$, the $\alpha$-particle, and the tritium
nucleus $t$. The nuclei $\alpha$ and $t$ are not treated as
structureless particles; their wave functions are constructed on
the nucleonic level and only nucleon motion within the clusters is
approximated by simple shell-model configurations. The many-body
state then describes the correlated relative motion of the
different clusters in a fully anti-symmetrized wave function that
obeys the Pauli principle and thus also accounts for correlated
motion of the halo neutrons. The nucleon-nucleon interactions are
chosen to reproduce, {\it e.g.}, phase shifts in {\it NN},
$\alpha${\it N} and $\alpha\alpha$ scattering, and deuteron size
and binding energy. Additional effective nucleon-nucleon interactions are included to account for three-nucleon
interactions. This model clearly shows the best agreement with our experiment ($\triangledown$ in
Fig.\,\ref{fig:fig2}). Calculations for $^{11}$Li were performed both with and without possible excitations of
the $^9$Li core by the halo neutrons. Including these intrinsic excitations results in $r_{\rm c}(^{11}{\rm Li})
= 2.52$ fm, in good agreement with experiment, while neglecting them results in the much smaller value $r_{\rm
c}(^{11}{\rm Li}) = 2.28$ fm ($\blacktriangledown$ in Fig.\,\ref{fig:fig2}). Thus, within the framework of SVMC,
neutron correlations alone cannot reproduce the large change in $r_{\rm c}$ between $^9$Li and $^{11}$Li
observed in the experiment. The calculations rather indicate that the core is indeed perturbed and that this
perturbation accounts for most of the charge radius increase.
It will be interesting to see whether the model is also able to describe correlations in the momentum distributions of breakup fragments \cite{Sim99}. We also note that while the SVMC model clearly shows the best agreement with our measured nuclear charge radii, it's predictions for nuclear electromagnetic moments \cite{Var02} still have significant discrepancies from experimental values, indicating that further work is still needed.

\begin{acknowledgments}
This work is supported from BMBF contract No. 06TU203. Support
from the U.S. DOE Office of Science (B.A.B.), NRC through TRIUMF,
NSERC and SHARCnet (G.W.F.D. and Z.-C.Y.) is acknowledged. A.W.
was supported by a Marie-Curie Fellowship of the European
Community Programme IHP under contract number HPMT-CT-2000-00197.
We thank the target laboratory at GSI for providing the carbon
foil catcher, Nikolaus Kurz, Mohammad Al-Turany (GSI) and the ISAC
computer division at TRIUMF for support in data acquisition,
Melvin Good for help during installation of the experiment at
TRIUMF, and Ren\'e Roy for providing a liquid scintillator.
\end{acknowledgments}


\begin{thebibliography}{52}
\bibitem{Jen04}
A.~S.~Jensen {\it et al.}, Rev. Mod. Phys. {\bf 76}, 215 (2004).
\bibitem{Tan85}
I.~Tanihata {\it et al.}, Phys. Rev. Lett. {\bf 55}, 2676 (1985).
\bibitem{Wan04}
L.-B.~Wang {\it et al.}, Phys. Rev. Lett. {\bf 93}, 142501 (2004).
\bibitem{Pie01}
S.~C.~Pieper and R.~B.~Wiringa, Annual Rev. Nucl. Part. Science {\bf
51}, 53 (2001).
\bibitem{Bla92}
B.~Blank {\it et al.}, Z. Phys. A {\bf 343}, 375 (1992).
\bibitem{Yan00}
Z.-C.~Yan and G.~W.~F.~Drake, Phys. Rev. A {\bf 61}, 022504 (2000).
\bibitem{Yan03}
Z.-C.~Yan and G.~W.~F.~Drake, Phys. Rev. Lett. {\bf 91}, 113004 (2003).
\bibitem{Ewa04}
G.~Ewald {\it et al.}, Phys. Rev. Lett. {\bf 93}, 113002 (2004).
\bibitem{Bus03}
B.~A.~Bushaw {\it et al.}, Phys. Rev. Lett. {\bf 91}, 043004 (2003).
\bibitem{Bac05}
C.~Bachelet {\it et al.}, Eur. Phys. J. A {\bf 25 Supp. 1}, 31 (2005).
\bibitem{Jag74}
C.~W.~de Jager, H.~deVries, and C.~deVries, At. Data Nucl. Data
Tables {\bf 14}, 479 (1974).
\bibitem{Sic03}
I.~Sick, Phys. Rev. Lett. B {\bf 576}, 62 (2003).
\bibitem{Kop97}
S.~Kopecky {\it et al.}, Phys. Rev. C {\bf 56}, 2229 (1997).
\bibitem{Fri97}
J.~L.~Friar, J.~Martorell, and D.~W.~L.~Sprung, Phys. Rev. A {\bf
56}, 4579 (1997).
\bibitem{Pie02}
S.C.~Pieper, K.~Varga and R.B.~Wiringa, Phys. Rev. C {\bf 66},
044310 (2002).
\bibitem{Var95}
K.~Varga, Y.~Suzuki, and I.~Tanihata, Phys. Rev. C {\bf 52}, 3013
(1995).
\bibitem{Var02}
K.~Varga, Y.~Suzuki, and R.~G.~Lovas, Phys. Rev. C {\bf 66}, 041302
(2002).
\bibitem{Nef05}
T.~Neff, H.~Feldmeier, and R.~Roth, in 21. Winter workshop on
Nuclear Dynamics, Breckenridge, Colorado, USA, (2005).
\bibitem{Tom01}
M.~Tomaselli {\it et al.}, Nucl. Phys. A {\bf 690}, 298c (2001).
\bibitem{Nav03}
P.~Navr\'atil and W.~E.~Ormand, Phys. Rev. C {\bf 68}, 034305
(2003).
\bibitem{Nav98}
P.~Navr\'atil and B.~R.~Barrett, Phys. Rev. C {\bf 57}, 3119 (1998).
\bibitem{Zhu93}
M.~V.~Zhukov {\it et al.}, Phys. Rep. {\bf 231}, 151 (1993).
\bibitem{For02}
C.~Forss\'en, V.~D.~Efros, and M.~V.~Zhukov, Nucl. Phys. A {\bf
706}, 48 (2002).
\bibitem{Ege02}
P.~Egelhof {\it et al.}, Eur. Phys. J. A {\rm 15} 27 (2002).
\bibitem{Zhu05}
M.~V.~Zhukov, Chalmers University of Technology, (2005).
\bibitem{Las62}
K.~E.~Lassila {\it et al.}, Phys. Rev. {\bf 126}, 881 (1962).
\bibitem{Sto93}
V.~G.~J.~Stoks {\it et al.}, Phys. Rev. C {\bf 48}, 792 (1993).
\bibitem{Nef05b}
T.~Neff, H.~Feldmeier, and R.~Roth, Nucl. Phys. A {\bf 752}, 321 (2005).
\bibitem{Sim99}
H.~Simon {\it et al.}, Phys. Rev. Lett. {\bf 83}, 496 (1999).
\end{thebibliography}
\end{document}